\documentclass[aps,twocolumn,prl,preprintnumbers,amsmath,amssymb,superscriptaddress]{revtex4-1}

\usepackage{graphicx}% Include figure files
\usepackage{bm}% bold math
\usepackage{hyperref}
\usepackage{float}
\usepackage{diagbox}
\usepackage{float}
\usepackage{multibib}
\usepackage{natbib}
\begin{document}

\title{Absence of cyclotron resonance in the anomalous metallic phase in InO$_x$}

\author{Youcheng Wang$^1$,   I. Tamir$^2$,  D. Shahar$^2$,  and N. P. Armitage$^1$\\
\medskip
$^1$ Institute for Quantum Matter, Department of Physics and Astronomy, Johns Hopkins University, Baltimore, Maryland 21218, USA\\
$^2$  Department of Condensed Matter Physics, Weizmann Institute of Science, Rehovot 76100, Israel}

\date{\today}

\begin{abstract}		
It is observed that many thin superconducting films with not too high disorder level (generally R$_N/\Box < 2000 \Omega$) placed in magnetic field show an anomalous metallic phase where the resistance is low but still finite as temperature goes to zero.  Here we report in weakly disordered amorphous InO$_x$ thin films, that this anomalous metal phase possesses no cyclotron resonance and hence non-Drude electrodynamics. The absence of a finite frequency resonant mode can be associated with a vanishing downstream component of the vortex current parallel to the supercurrent and an emergent particle-hole symmetry of this metal, which establishes its non-Fermi liquid character.
\end{abstract} 
\maketitle

Conventional wisdom dictates that electrons confined to two dimensions and cooled to the limit of zero temperature can only be superconducting or insulating. Yet a number of examples of zero temperature 2D metallic states exist. A particularly interesting example is the anomalous metal where thin superconducting films in magnetic field may exhibit a saturated, but non-zero resistance much smaller than the normal state at low temperature.  This is a widely observed phenomenon that manifests in a diverse number of physical systems:  disordered thin films, \cite{Jaeger89a,yazdani95a,ephron96a,mason99a, Hebard1992, Chervenak00a,lin12a, liu2013microwave}, Josephson junctions arrays \cite{van1992field, han2014collapse}, artificially patterned superconducting islands \cite{eley2012approaching} and interfacial superconductivity \cite{schneider2009electrostatically}.  It is under appreciated how anomalous this otherwise simple behavior is. The natural behavior of bosons in the limit of zero temperature is either condensed or localized.  But here, due to the low resistance of the sample and the obvious manifestation of the superconducting transition one requires the presence of Cooper pairs that move diffusively even in the apparent limit of zero temperature. In our below work we assume that this anomalous metal phase has been established by these previous experiments.   

Although observed ubiquitously the nature of this anomalous metal is still unclear \cite{das99a, mason99a,kapitulnik01a, phillips02a,galitski05a,davison2016hydrodynamic,mulligan2016composite, spivak2008PRB}.  It was observed in previous experiments \cite{liu2013microwave} that its optical response is characterized by a very narrow conductance peak with a width that is orders of magnitude less than the normal state scattering rate (which is typically $\sim$ 100 THz).  It was shown that over short time scales, the system possesses a frequency dependent phase stiffness \cite{liu2013microwave}, indicating that superconducting correlations are retained on small time and length scales.  But how this is possible in a zero temperature dissipative phase, and to what extent this state's phenomenology is different from conventional metals is unclear.
   
In this work we have performed broadband microwave measurements of frequency, temperature, and field dependence of the complex microwave conductance on a low-disorder 2D superconducting InO$_x$ film. These measurements have been extended to the very low field regime when samples are still highly conductive, giving unprecedented insight into how this anomalous metallic state develops from superconductivity. We show that even at small magnetic fields, although the superconducting delta function is retained, the low frequency dissipative modes almost immediately form. With further applied field the delta function is suppressed leaving only the dissipative peak. Among other aspects we demonstrate that although on some level this anomalous metal can be characterized as a high mobility metal, the system posses $no$ cyclotron resonance, which is a ubiquitous feature of high mobility metals with conventional electrons. Our observation has much in common with recent observations that the anomalous metal state has -- over a range of fields -- no Hall effect \cite{Breznay17a}. Taken together, these works show that the anomalous metal state has an emergent particle-hole symmetry, that although was previously considered to be a property of the SIT \cite{SIT} quantum critical point \cite{Hebard1992,fisher1991hall}, is exhibited here over an entire intervening phase.

\begin{figure}
    \centering
	\includegraphics[width=0.35\textwidth]{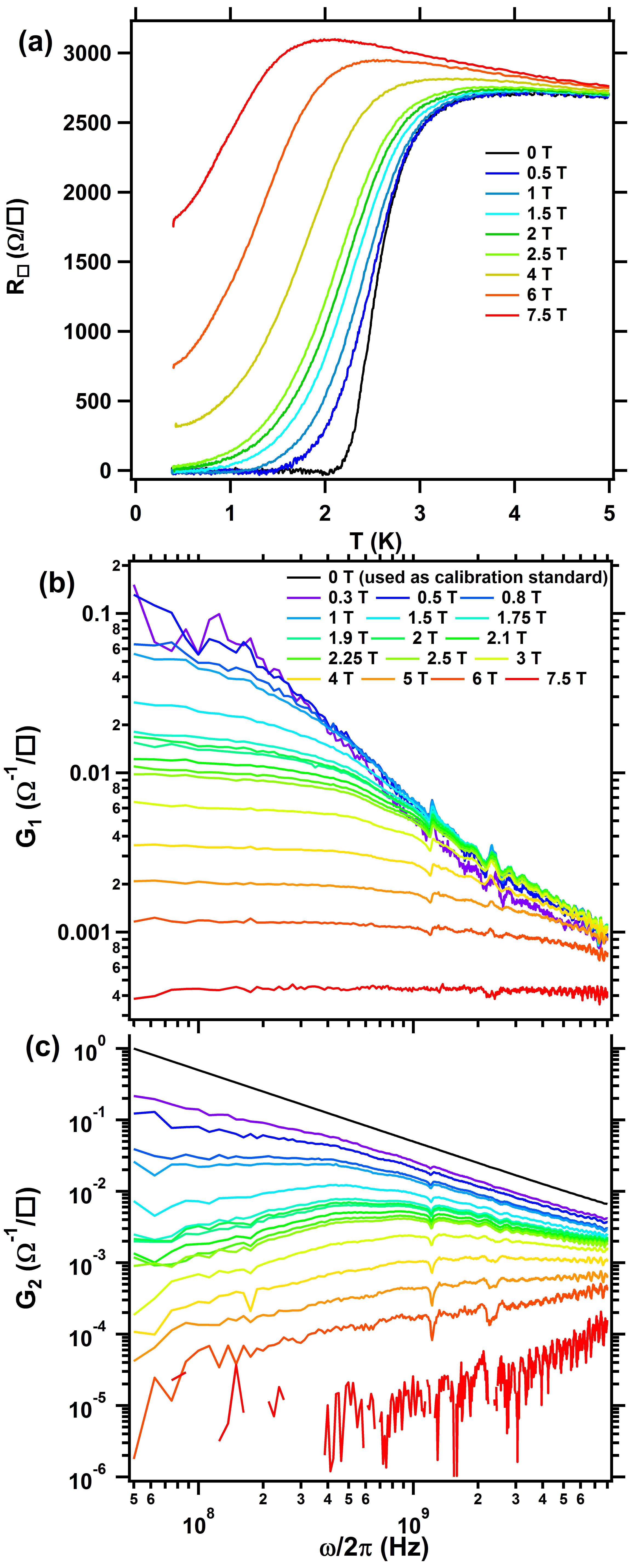}
	\caption{(color online) (a) Temperature dependence of the DC sheet resistance under different magnetic fields. (b,c) Complex sheet conductance at base temperature on a log-log scale. (b) Real part $G_1$. $G_1$ ($H$ = 0 T) is not plotted, but can be found in Fig. \ref{fig:Fig2}. (c) The corresponding imaginary part.}
	\label{fig:Fig1}
\end{figure}
 
In Fig. \ref{fig:Fig1}(a), we show two terminal DC sheet resistance R/$\Box$ measured in the Corbino geometry as function of temperature. The resistance as a function of field at the low temperature of $\approx$ 0.4 K is presented in Supplemental Information (SI). One can see the characteristic phenomenon of the anomalous metallic phase \cite{reentrance} with apparently finite resistances persisting to the low temperature limit. The ``mean field" superconducting transition temperature $T_{c0}$ scale at zero field is estimated to be about 2.5 K.  As the magnetic field increases, the low temperature ($\ll T_{c0}$) resistance departs from the zero field curve and at base temperature (0.38 K) first becomes distinguishable from zero above $\approx$ 2 T (see SI) which gives us an estimate for the critical field for the superconductor to metal transition $H_{sm}$. For fields above 2 T, the temperature dependence is weak as $T \rightarrow 0$ and the resistance appears to saturate at finite values much smaller than the normal state.  For 4, 6 and 7.5 T, a weak insulating dependence develops around $T_{c0}$ but the resistance still decreases and then levels out at low T. It is important to note that this is likely only a true phase transition in the limit of zero temperature and that any 2D superconductor in magnetic field is expected to have a small (but possibly undetectable resistance) at finite temperature.

In Figs. \ref{fig:Fig1}(b)-(c), we show the real ($G_{1}$) and imaginary ($G_{2}$) parts of the complex conductance at the base temperature at different fields on log-log plots.  This data was measured in a novel broad band Corbino spectrometer \cite{Corbino} down to 0.38 K. Zero field data was used as a superconducting short calibration standard by assuming the imaginary part is ideal (for details see SI). At zero field, the real part has presumably the response of a Dirac delta function given by $ G_{1}(\omega) = \frac{\pi n_s e^2}{2m}\delta(\omega), $ the spectral weight (e.g. the integrated area) of which corresponds to superfluid density $\rho_s $. By Kramers-Kronig consistency, the imaginary part $G_{2}(\omega) =\frac{n_s e^2}{m\omega} =\frac{2  e^2}{ \pi \hbar}\frac{k_B T_{\theta}}{ \hbar \omega}.$

Here one can rewrite the spectral response in terms of a phase stiffness $T_{\theta}$, which is the energy scale to put a twist in the phase of the superconducting order parameter. With the presence of small external field, the real part immediately obtains a finite value at finite frequency alongside the delta function, which is retained over some range in field (the latter inferred from the vanishing dc resistance).  Consistent with this, the imaginary conductance deviates from a $1/\omega$ dependence at low frequencies which can be interpreted as suppressed long-range phase coherence. The high frequency response near 8 GHz retains its $1/\omega$ dependence but with reduced magnitude. For fields just below 1 T, $G_{2}$ is flat over almost a decade of frequencies which further suggests that multiple conductance channels exist and a single Drude term or a $1/\omega $ dependence is insufficient to account for the low frequency response. Above $H_{sm}$, the imaginary part develops a broad maximum that was previously interpreted to indicate the fluctuation frequency $\Omega$ \cite{liu2013microwave} on the approach to the superconductor-metal transition.
 
\begin{figure*}
	\centering
	\includegraphics[width=0.9 \textwidth]{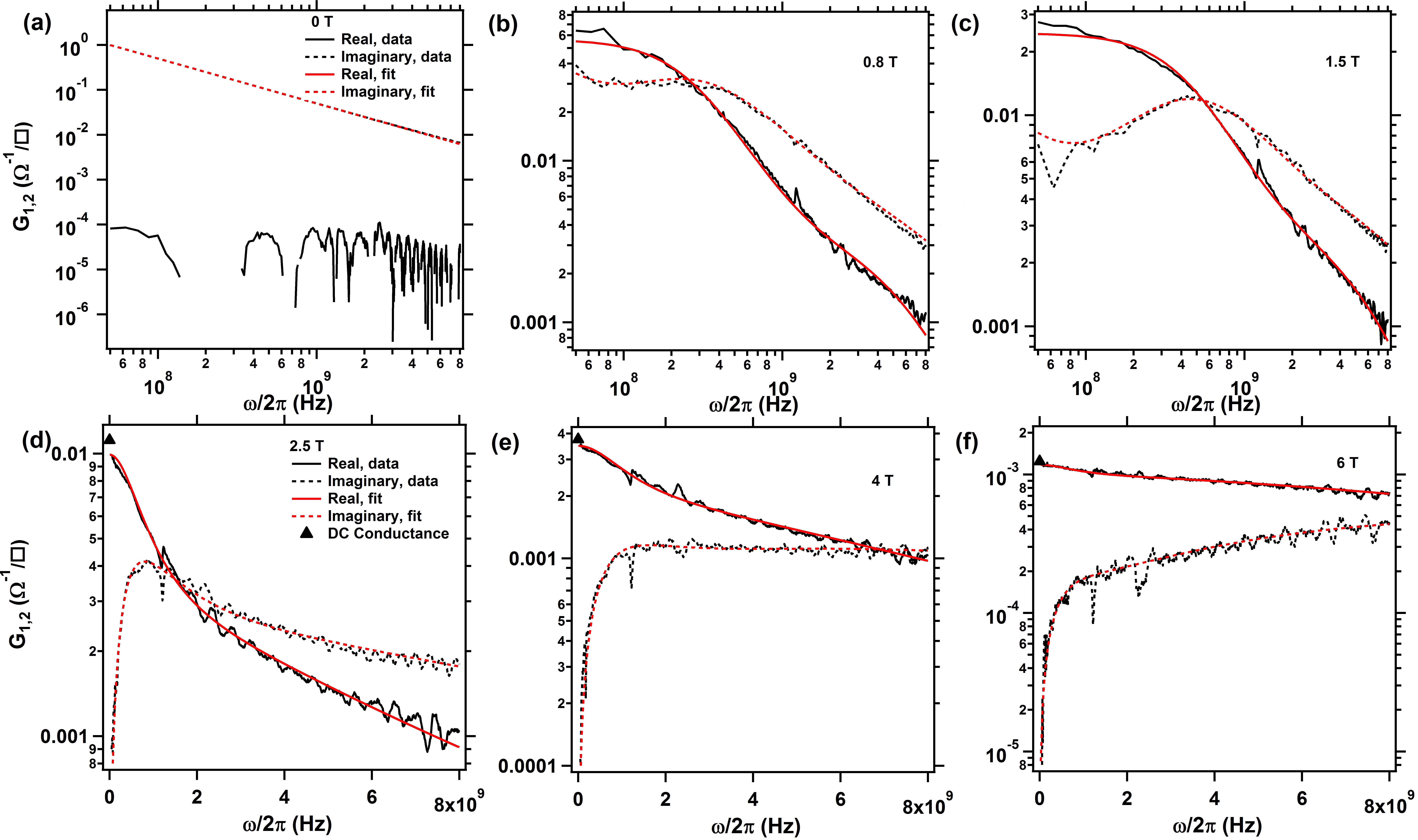}
	\caption{(color online) Complex conductance fits for (a) 0 T, (b) 0.8 T, (c) 1.5 T, (d) 2.5 T, (e) 4 T, and (f) 6T.  (a)-(c) are on log-log scales while (d)-(e) only have the vertical axis on log scale to show DC data.  The black triangle represents the co-measured DC data.  Models used in the fitting are described in the text.}
	\label{fig:Fig2}
\end{figure*}

Notably, as the narrow low frequency peak in $G_1$ decreases in magnitude and broadens, to within experimental uncertainty, it does not shift from zero frequency.  If the dissipative state above $H_{sm}$ were a conventional high mobility metal, one would expect to observe classical cyclotron resonance whereby a finite frequency peak will be exhibited in $G_1$.   Cyclotron resonance in which charges undergo periodic orbits in magnetic field is an essential property of conventional metals.  Classically this resonance frequency is expected to be $\omega_c = eB/m$.  In a Galilean invariant system, via Kohn's theorem \cite{kohn1961cyclotron} this frequency is expected to be unchanged by interactions with an inferred mass $m$ equal to the bare electron mass ($m_e$) itself.  In low density 2DEGs with Fermi wavelengths much larger than the lattice constant, an approximate Galilean invariance is obtained and $m$ is found to be the band mass \cite{abstreiter1974cyclotron, kennedy1975frequency}. From a minimal fitting of the complex conductance using the expression for semi-classical transport \cite{fitting} we can set an upper bound on the cyclotron resonance to be less than 65 MHz at 6 T. This is an exceedingly small number and in the context of semi-classical transport implies an effective mass greater than 2,500 $m_e$ or internal effective field that is 1/2,500 of the applied field!  Therefore we ascertain that to within our experimental uncertainty the anomalous metal state has no cyclotron resonance.

The spectra can in principle be fit to a combination of three phenomenologically assigned terms: a zero frequency delta function, a finite width peak centered at zero frequency that is characteristic of the anomalous metal, and a broader background  (See SI for a decomposition). In the fits these contributions are modeled as classical Lorentzians \cite{fitting}. Fig. \ref{fig:Fig2} shows the fitting at 6 characteristic fields. In the superconducting state (Figs. \ref{fig:Fig2}(a)-(c)), the imaginary conductance exhibits a diverging trend as frequency goes to zero. This observation can be related to the existence of the delta function in the real conductance.   For small, but finite field (say 0.8 T) the contribution of multiple features in the spectra is apparent.   In the superconducting regime, we fit the spectra to the three terms and in the anomalous metal, only two.  In the superconducting state, the presence of very small, but still finite width to the ``delta" function contribution (say due to finite temperature) does not effect these fits as any width is below the low frequency end of the spectrometer and indistinguishable from a delta function.  In the anomalous metal, the delta function is obviously absent and the imaginary part extrapolates to zero at zero frequency. However, the conductance has non-Drude lineshape, as demonstrated in Fig. \ref{fig:Fig2}(e) for 4 T where the imaginary part is almost flat from 1 GHz to 8 GHz.   This data fits well with a Drude term of the width of $\sim$ 8 GHz and a much narrower term ($\sim$ 1 GHz wide) which compensates the imaginary part at higher frequencies.  In the metallic regime, it is not clear if one should interpret the spectra as truly the sum of two channels, or if the transport is single channel and the fits should be considered only phenomenological.

However, given these fits it is interesting to see how the spectral weight of the various features evolves as a function of field.  In phase  models, the spectral weight can be identified with a phase stiffness ($T_{\theta}$), which one finds by multiplying the plasma frequency (explained in the SI) squared by $\frac{\hbar}{2k_{B} G_Q}$ (with the quantum of conductance $G_Q \equiv 4e^2/h$) to get the stiffness in the units of degrees Kelvin. In Fig. \ref{fig:Fig3}, we plot the spectral weight of the delta function, the low frequency dissipative conductance, and the total as a function of field. The spectral weight of the low frequency dissipative conductance was obtained by integrating it up to 10 GHz.

With small applied field the spectral weight of the delta function falls extremely quickly with field.   Although part of the spectral weight is transferred to the low frequency peak, the rest is not and presumably goes to energy scales of order the normal state scattering.   By the time the metallic state is being approached near 2 T, the delta function spectral weight has been suppressed to very small values. The transition to anomalous metal state occurs by suppressing the delta function altogether and leaving behind the anomalous low frequency peak as the dominant conducting channel at low $\omega$.

Our observation of no cyclotron resonance is consistent with recent transport experiments by Breznay and Kapitulnik on InO$_x$ and TaN$_x$ thin films \cite{Breznay17a}.  They report that the Hall resistance of  InO$_x$ is indistinguishable from zero upon entering the anomalous metal regime at $ \approx $ 2.0 T and remains indistinguishable from zero up to a higher field scale that is still within the anomalous metal region.   We speculate that in the two component decomposition of our spectra, the narrow low frequency component exhibits no Hall response and the crossover to a regime with Hall response occurs when broader feature (which we associate with the normal state) starts to dominate the spectrum.  These results suggest that the anomalous metal state itself has a robust particle-hole symmetry.   Previously it was suggested in Ref. \cite{fisher1991hall} that the SIT quantum critical point had a emergent particle-hole symmetry that ensured that the pseudo-Lorentz force was zero at the transition giving zero Hall effect and a vanishing cyclotron resonance.  Working from the perspective that here the critical point of the SIT broadens into a \textit{phase}, it appears that the anomalous metal exhibits a similar symmetry.   We should point out that a particle-hole symmetry as such \textit{is not} conventionally a property of superconductors themselves.   In the flux-flow regime, superconductors show a Hall response \cite{hagen1990anomalous,nozieres1966motion,jing1990flux} that depends on the details of the vortex damping.   Even in the fluctuation regime above T$_c$ particle-hole symmetry appears to be broken by terms that depend on the derivative of the density of states with energy \cite{fukuyama1971fluctuation,BreznayHall}, giving the Aslamazov-Larkin superconducting fluctuation contribution a finite Hall effect. However, note that although a corresponding cyclotron resonance-like energy scale can be defined in such calculations, it is expected to be \textit{larger} than the normal metal cyclotron resonance by a factor of $k_F l$ \cite{michaeli2012hall} and moreover does not necessarily result in a resonance \cite{Finkelstein18a}. Therefore the particle-hole symmetry encountered here requires further perspective.  It would be interesting to consider older theories  \cite{das99a, phillips02a, galitski05a, spivak2008PRB} in the context of these newer experiments.

\begin{figure}
	\centering
	\includegraphics[width=0.35 \textwidth]{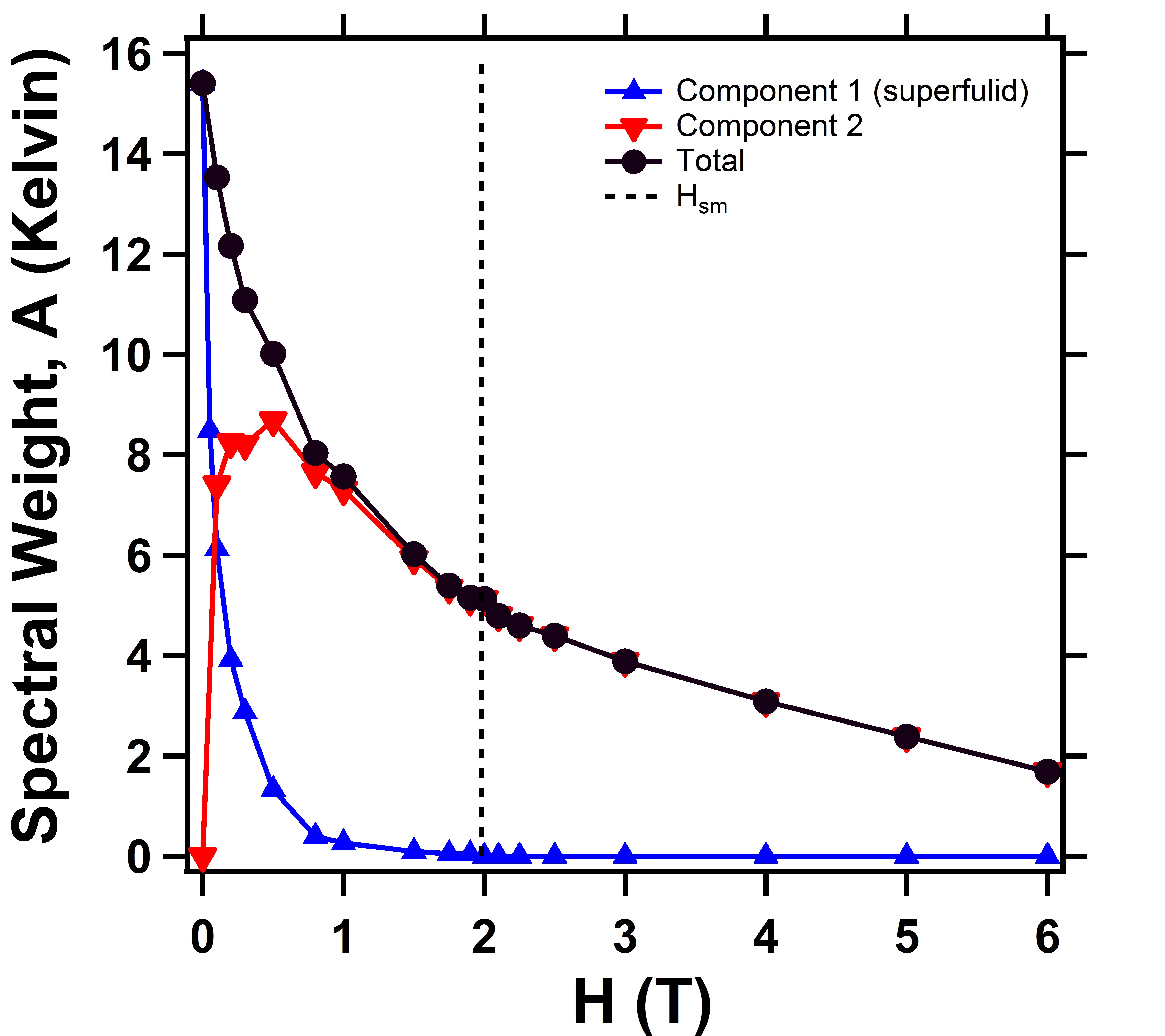}
	\caption{Spectral weight from fits in the units of Kelvin for the delta function and Drude terms. Blue and red corresponds to the superfluid (delta function) and the slowly decaying component respectively. The sum of the total spectral weight is in black. The spectral weight for component 2 at 0.05 T is not plotted because the real part of conductance falls below sensitivity. See Sec VI. in SI for details.}
	\label{fig:Fig3}
\end{figure}

Our observations are in-part reminiscent of the phenomenology of composite fermions of half-filled Landau levels in 2D electron gasses.  In such systems a cyclotron resonance mode is found that depends not on the applied magnetic field, but instead an effective magnetic field $B_{eff}$ which is the physical applied field minus a Chern-Simons field \cite{halperin2004fermion}. The cyclotron resonance disappears at fields that correspond to even-denominator Landau level filling fractions \cite{kukushkin2002cyclotron}, however the actual physical Hall effect remains finite. In contrast, in the present case the cyclotron resonance is absent over a range of fields and the Hall effect vanishes. Connection of the theory of composite fermions to the SIT quantum critical point and intervening metals have been recently made \cite{mulligan2016composite}. We point out in this regard that the density of charges at the transition is of order the number of vortices (see SI).

Recently, Davison et.al. employed a hydrodynamic approach and memory matrix formalism to calculate the dynamical conductivity of a phase fluctuating superconductor in the incoherent limit  \cite{davison2016hydrodynamic}. With strong time-reversal and parity symmetry breaking, the dynamical conductivity is allowed to have a \textit{supercyclotron} resonance mode at $\omega^{\star} = \Omega^H -i \Omega$.  This mode is allowed if the vortex current $\bm{J_v} \equiv n_v q_v \bm{v_v} $ has a downstream component with the supercurrent $\bm{J_s} \equiv \rho_{s} \bm{v_s}$, other than a transverse component due to the superfluid Magnus force \cite{nozieres1966motion}, i.e. $\bm{F_M} = -q_v \rho_s (\bm{v_s}-\bm{v_v}) \times \hat{\bm{z}} \  \Phi_0$  for a single vortex with vorticity $q_v$ in S.I. unit. The downstream vortex current then generates an emergent transverse electric field that bends the path of supercurrent.  Our data constrains $ \Omega^H$  to be less than $\sim$ 65 MHz, which implies that the downstream component is much smaller than the transverse component that dominates vortex dynamics. 

In this work, we have investigated the low frequency dynamical conductivity of a thin superconducting film in vicinity of the superconductor to metal transition.  Remarkably  the system posses $no$ cyclotron resonance, which is a ubiquitous feature of high mobility metals composed of conventional electrons.   Our observation taken with that the anomalous metal state has -- over a range of fields -- no Hall effect \cite{Breznay17a} shows an emergent particle-hole symmetry.  This was previously considered to be a property of the SIT quantum critical point, but here is exhibited over an entire intervening phase.  The anomalous metal should be considered a unique state of matter, which cannot be understood in terms of the conventional theory of metals. 

This work at JHU was supported by NSF DMR-1508645. Work at the Weizmann Institute was supported by the Israel Science Foundation(ISF Grant no. 751/13) and The United States-Israel Binational Science Foundation (BSF Grant no. 2012210). We would like to acknowledge A. Finkelstein, S. Hartnoll,  A. Kapituilnik, S. Kivelson, M. Mulligan, S. Parameswaran, P. Phillips, and S. Raghu for helpful conversations.

\onecolumngrid
\newpage
\setcounter{figure}{0}  
\renewcommand\thefigure{S\arabic{figure}} 
\renewcommand{\figurename}{Fig.} 

\renewcommand{\author}{}
\renewcommand{\title}{Absence of cyclotron resonance in the anomalous metallic phase in InO$_x$}
\begin{center}
	\textbf{\large Absence of cyclotron resonance in the anomalous metallic phase in InO$_x$}
\end{center}
\begin{center}
	\author{Youcheng Wang$^1$,   I. Tamir$^2$,  D. Shahar$^2$,  and N. P. Armitage$^1$\\
		\medskip
		$^1$ Institute for Quantum Matter, Department of Physics and Astronomy, Johns Hopkins University, Baltimore, Maryland 21218, USA\\
		$^2$  Department of Condensed Matter Physics, Weizmann Institute of Science, Rehovot 76100, Israel}
\end{center}

\section{Methods}
\label{Methods}

Broadband microwave experiments were performed in a home-built Corbino microwave spectrometer coupled into a He-3 cryostat \cite{liu2011dynamical, liu2013microwave}.  Such a system can measure the complex conductance over a broad spectral range from 50 MHz to 10 GHz down to temperatures of $\sim$ 360 mK.  Samples terminate a coaxial waveguide by being press fit into a custom Corbino shaped connection as described in Refs. \cite{scheffler05b,liu2014broadband}. To ensure a good electrical connection a Cr(10 nm)/Au(200 nm) metal layer in the shape of a ``donut" pattern was evaporated onto silicon substrates in advance on InO$_x$ deposition.  Then another Au(200 nm) layer using the same mask was added on top.  We measure the complex reflectivity of the thin film using a network analyzer (see Supp. Fig. \ref{fig:Fig1}), from which complex sheet impedance and conductance can be obtained. Two terminal DC resistance can be simultaneously  measured $via$ a bias tee.  The DC resistance without microwave illumination was used to carefully check and correct for any microwave induced heating. To calibrate the system (particularly for low fields) and remove the effects of loss and extraneous reflections in the transmission line a multi-step calibration procedure was followed. First, three calibration samples with known reflection coefficients were measured (short, open and load) \cite{BoothPRL96a,LeePRL01b,scheffler05b, KitanoPRB09a,  liu2011dynamical, liu2013microwave}.  Calibrations were performed at 0, 0.5, 2, 4, 6, and 7.5 T and interpolated in between unless otherwise specified. Next, we ``recalibrate", by first assuming at 0 T the sample's imaginary impedance is ideal given by a linear fit to the initially calibrated imaginary part of the impedance and then using the 0 T superconductor as a new short calibration.  This procedure allows us to get accurate conductance data even deep into the superconducting state when the sample's is much less that 50 $\Omega$.  After performing a small further correction due to the effects of the substrate, (see SI Sec. \ref{Cal}) the complex conductance of the InO$_x$ film can be isolated at all relevant fields and temperatures.

High-purity (99.999 \%) In$_2$O$_3$ was e-gun evaporated in high vacuum environment onto clean high-resistivity Si substrates quenched at room temperatures.  The thickness was monitored in-situ by a quartz crystal resonator to approximately 30 nm.  TEM-diffraction patterns of InO$_x$ films prepared in the same fashion show diffusion rings with no diffraction spots. The lack of the reentrant behavior in $R$ versus $T$ curves\cite{MurthyPRL04a,CranePRB07b} further support that the films are morphologically homogeneous.  InO$_x$ has been used extensively recently in investigations of the superconductor-insulator transition (SIT)\cite{MurthyPRL04a,MurthyPRL05a,Murthy06a,Steiner05a,CranePRB07a,Gantmakher01a} and our samples show behavior in broad agreement with previous work.

\section{Calibrations}
\label{Cal}

\begin{figure}
	\centering
	\includegraphics[width=0.5\textwidth]{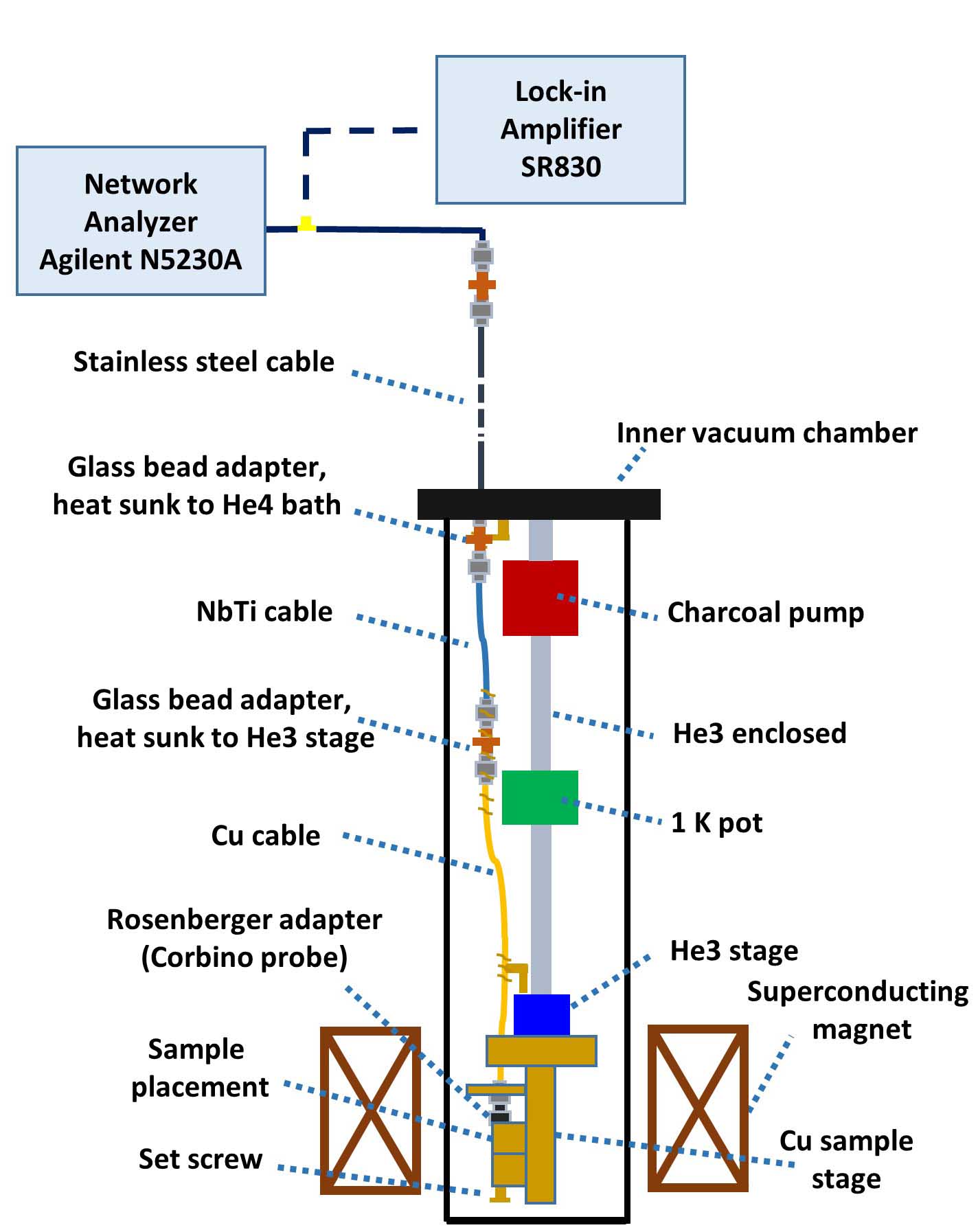}
	\caption{Schematic of the broadband Corbino microwave experiment setup.}
	\label{fig:Fig1}
\end{figure}

Our home-built broadband microwave ``Corbino" spectrometer (Fig.\ref{fig:Fig1}) uses a network analyzer to directly measure the complex reflection coefficient $S_{11}^{m}$ from a coaxial transmission line that is terminated by the sample.  To obtain the actual reflection coefficient $S_{11}^{a}$ from the sample surface itself, one needs to know three complex error coefficients \cite{scheffler2004broadband} $E_S$, $E_D$, and $E_R$, obtained from measurements on three calibration standards the response of which we presumably know $a \ priori$, namely a short ($S_{11}^{a} = -1$), an open ($S_{11}^{a} = 1$), and a load ($S_{11}^{a} = \frac{Z_L-Z_0}{Z_L+Z_0}$). $Z_0$ is the intrinsic impedance of coax cable which is  50 $\Omega$ for our experiment and $Z_L$ is the impedance of load. Then given the measured reflection coefficient from the sample $S_{11}^{m}$, the  actual reflection from the sample surface is \cite{liu2014broadband,liu2013broadband}:
\begin{equation}
S_{11}^{a} = \frac{S_{11}^{m}-E_D}{E_R+E_S(S_{11}^{m}-E_D)}.
\label{S11}
\end{equation}
By impedance matching, the effective sheet impedance of a sample surface follows from the measured reflection coefficient $S_{11}^{a}$
\begin{equation}
Z_{s}^{eff} = g \frac{1+S_{11}^{a}}{1-S_{11}^{a}}Z_0,
\label{Impd}
\end{equation}
where $g\equiv 2 \pi / ln(r_2/r_1)$ is the geometric factor for the Corbino measurements. For our sample, the outer and inner radius of the Corbino doughnut is 2.3 mm and 0.7 mm respectively and is defined by the mask used when evaporating contacts.  Assuming that the substrate impedance of the thin film sample is $Z_{s}^{sub}$, then the sample impedance $Z_{s}$ and measured effective impedance $Z_{s}^{eff}$  are related by 
\begin{equation}
Z_{s}^{eff} = \frac{Z_s}{1+Z_s/Z_{s}^{sub}}.
\label{subcorr}
\end{equation}
Hence a substrate correction (which will be elaborated on below) is necessary if $Z_s$ is comparable to $Z_{s}^{sub}$. In the thin film approximation, $Z_s =1/G \approx 1/\sigma d $ where $\sigma \equiv \sigma_{1} + i \sigma_2 $ is complex conductivity. See Ref. \cite{liu2013broadband,liu2011dynamical,BoothPRL96a} for more details. 

\begin{figure*}
	\centering
	\includegraphics[width=1\columnwidth]{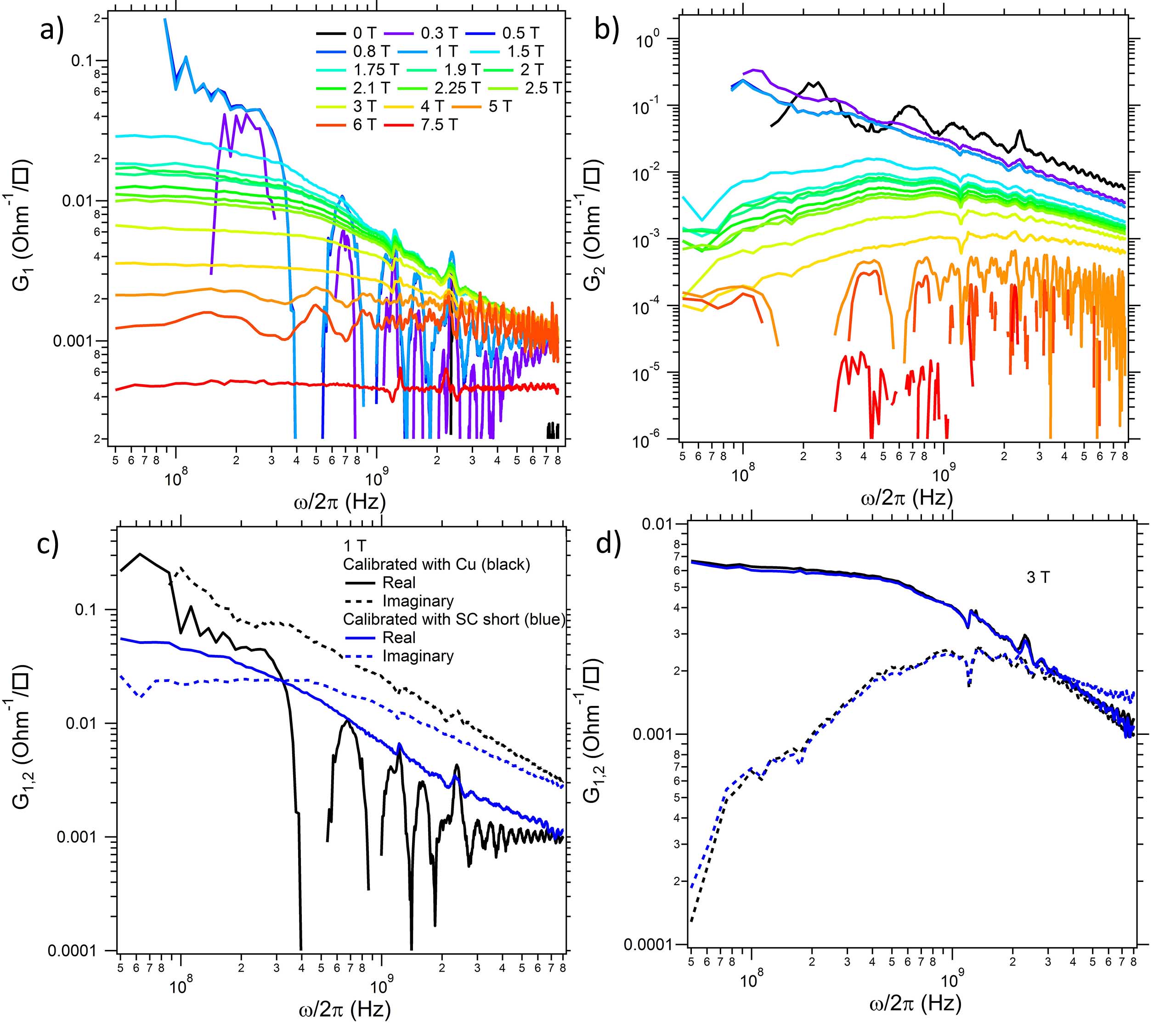}
	\caption{(color online)  (a),(b) Real and imaginary part sheet conductance of the sample at different fields from initial calibration using Cu as short and substrate correction. (c), (d) Comparison of the conductance at 1 T and 3 T between using Cu as short and the sample itself at zero field as the short.}
	\label{fig:Fig2}
\end{figure*}

As discussed in the main text, a modified calibration procedure from our previous works \cite{liu2014broadband,liu2013broadband} was used to increase the range of sample impedances that could be measured.   This was particularly useful to extend the range of  measurements to very low fields where sample impedances are very low.   Here we compare our previous scheme of calibrations to the present scheme.   One can see that substantial improvements are made.

In the conventional calibration, we used a bulk slab of Cu as the short, a bare intrinsic Si ($\rho >$ 10,000 $ \Omega\cdot cm$) wafer as an open, and a 20 nm NiCr film evaporated on Si as a load (impedance of 37 $\Omega$), to calibrate sample response at each frequency/temperature/magnetic field. All standards were measured at 0 T, 0.5 T, 2 T, 4 T, 6 T, and 7.5 T. The sample response at other fields were calibrated with error coefficients linearly interpolated between the measured set of fields.

The sample response in the normal state (at 6 K) was used to obtain the substrate impedance. Since the normal state scattering rate of the sample far exceeds the high frequency limit of our experiment (10 GHz), we assume that $Z_s$ at 6 K is purely real, and the real part of impedance is equal to DC resistance. The substrate impedance $Z_{s}^{sub}$ was calculated from Eq. \ref{subcorr} accordingly. We assumed that substrate impedance has little temperature dependence in the temperature range of 0.4 to $\sim$ 6 K considering the substrate is intrinsic silicon.  The effective conductance following from this procedure is presented in Supp. Fig. \ref{fig:Fig2}(a) and (b). Comparing it with the data displayed in the main text that used an additional step of calibration (discussed below), one finds that the conductance is much noisier (particularly at low fields) with a number of obvious artifacts.

As discussed in the main text, we can improve the calibration procedure by making an assumption that the sample at zero field and low temperature has the response of an ideal conductor and use it as a short.   After initial calibration using the copper standard, we fit the extracted quantity $\omega G_2 $ to a constant (see Eq.2 in the main text) and assumed that that the imaginary part of $G$ of the short standard is given by it.  As noted in the main text, the assumption that the imaginary part is $\propto 1/\omega$ is reasonable because after this calibration procedure, the real part conductance at 50 MHz is by 4 orders of magnitude smaller than that of imaginary part, showing self-consistency. In principle, one could perturb around this simple assumption and might be able to obtain a more consistent response but obviously the correction would be very small.  After this step of further calibration, the substrate correction was performed as described above.

After the ``recalibration" the data is of much higher quality.    Two representative fields below and above $H_{sm}$ are shown in Supp. Fig. \ref{fig:Fig2}(c) and (d) to show the difference between using Cu as the short standard, and the sample at zero field as the short.    For low fields where the impedance is low, the recalibration procedure significantly improves the data.   At higher fields, where the impedance is higher the improvement is minimal.

\section{Field Dependence of Resistance}
The field dependence of resistance at base temperature during a sweep-up is shown in Supp. Fig. \ref{fig:Fig3}. 
\begin{figure}
	\centering
	\includegraphics[width=0.5 \columnwidth]{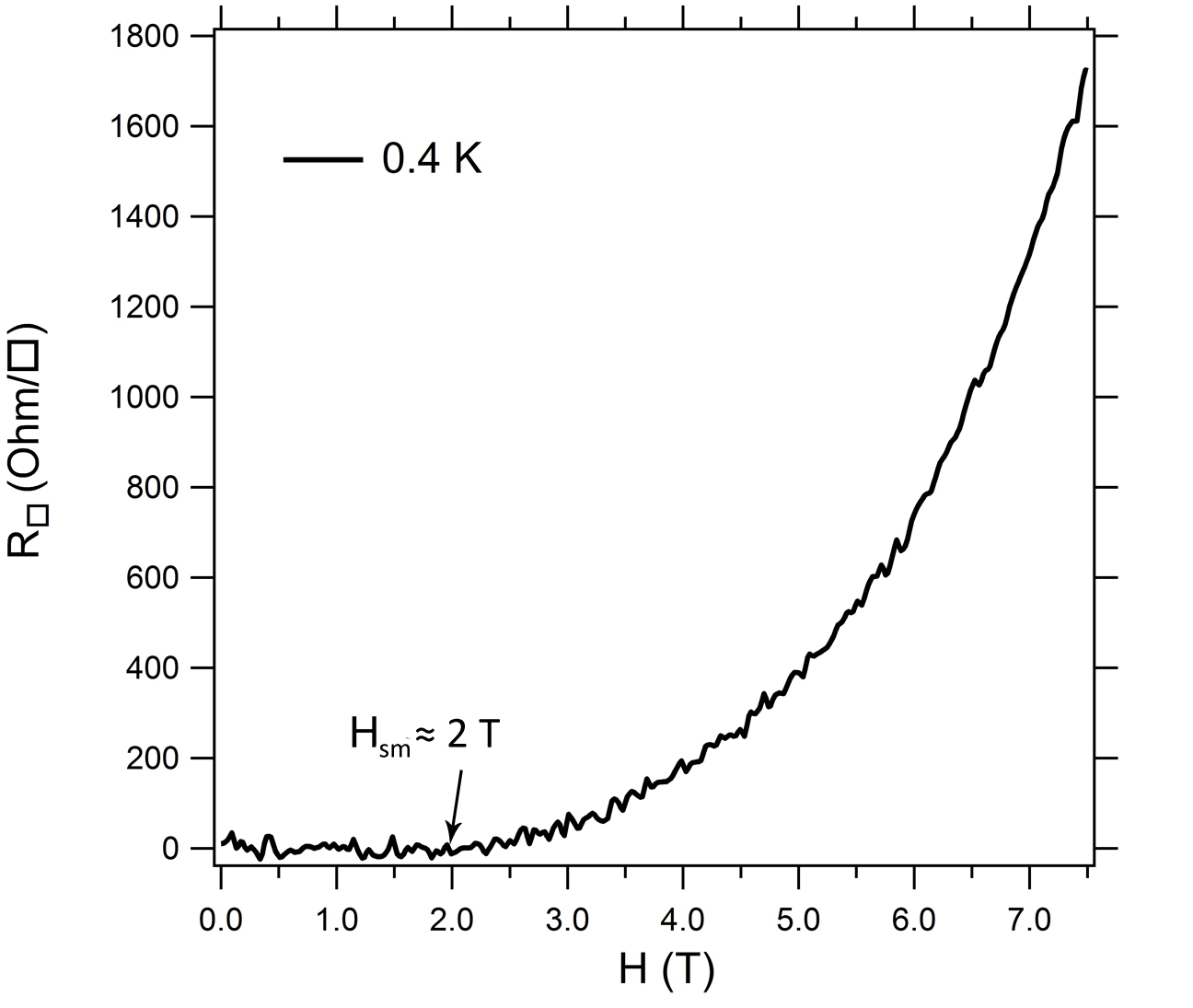}
	\caption{Sheet resistance as a function of perpendicular magnetic field during a sweep at $0.4 K$. Shown is the best estimate $H_{sm}$ for the field of the superconductor-metal transition.}   
	\label{fig:Fig3}
\end{figure}

\begin{figure*}
	\centering
	\includegraphics[width=1\columnwidth]{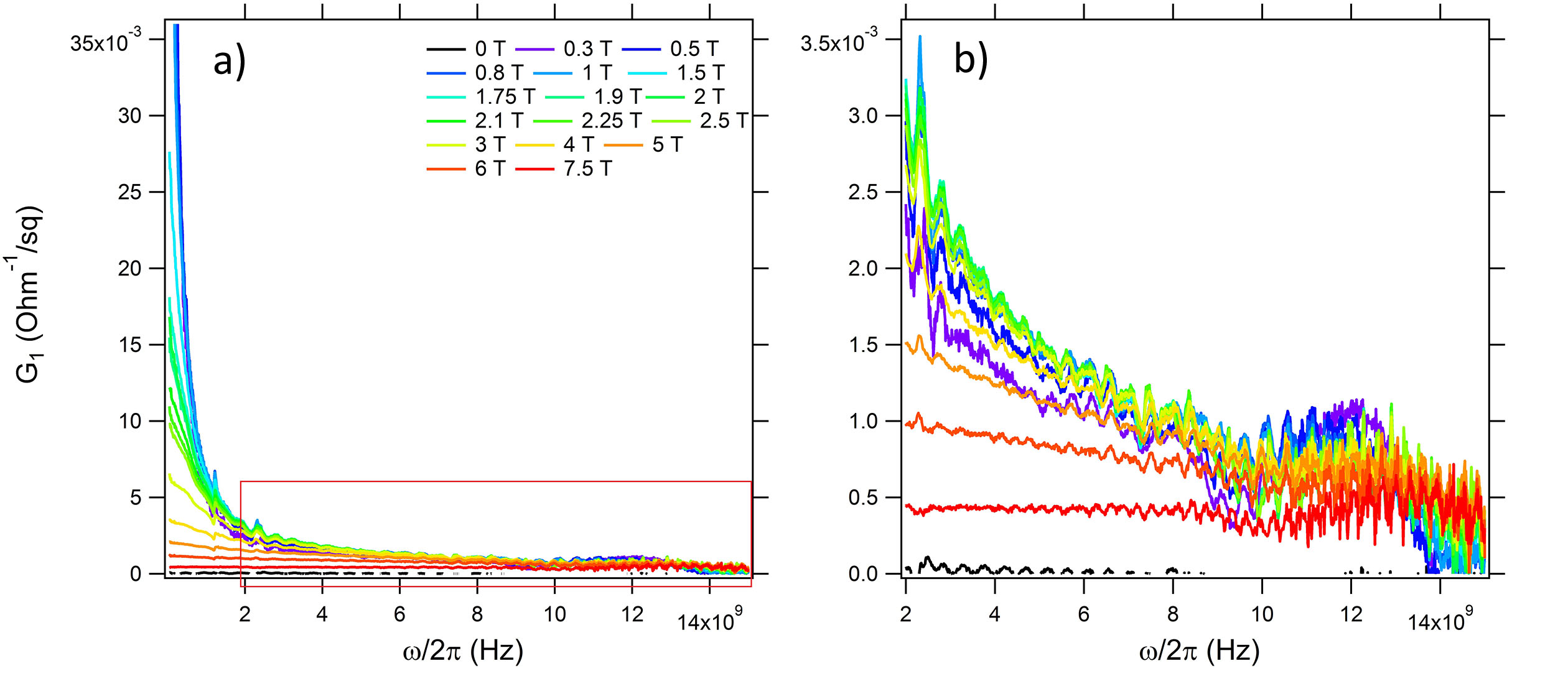}
	\caption{(color online)  (a) Field dependence of the real part of conductance on linear axes. (b) The zoomed-in graph of the area marked by a red rectangle in (a), which explicitly shows the high frequency range of the spectra.}   
	\label{fig:Fig4}
\end{figure*}

\section{Real Conductance on Linear Scale}

Here we show $G_1$ (Supp. Fig. \ref{fig:Fig4}) on linear scale, as a supplement to Fig. 3 in the main text of the paper.  Supp. Fig. \ref{fig:Fig4}(b) is an expanded view of the area marked by a red rectangle in Supp. Fig. \ref{fig:Fig4}(a). One can find that for fields 1.5 - 4 T, the spectra almost fell on top of each other at high frequency showing that the changes to the spectra with applied field are mostly a low frequency phenomenon.  The glitch at around 12 GHz comes from a significant standing wave resonance in the transmission line that could not be fully removed by calibration and exists for all samples we have measured.
\begin{figure}
	\centering
	\includegraphics[width=0.45 \textwidth]{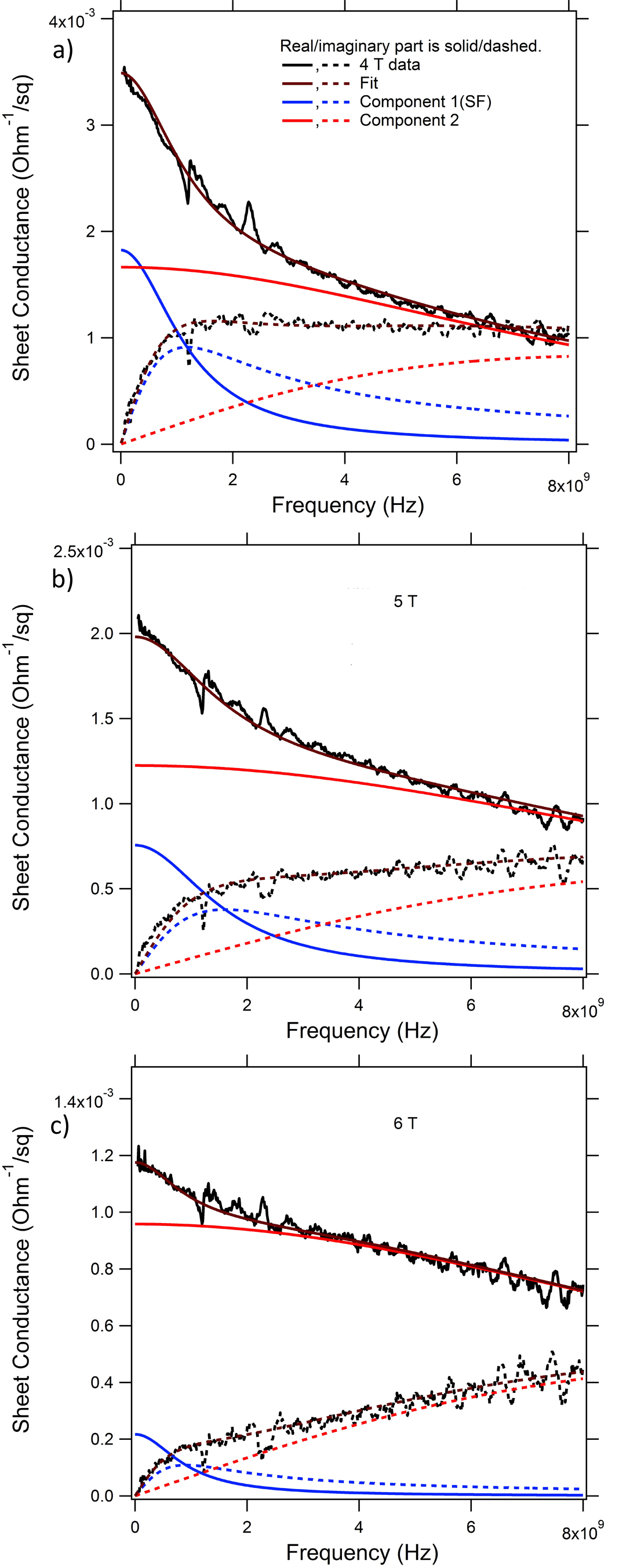}
	\caption{(color online)  The fitted curves for conductance at a) 4 T, b) 5 T, and c) 6 T. Two Drude components are separately plotted.}
	\label{fig:Fig5}
\end{figure}

\section{Fitting Models and Parameters}
Since the data is presented as sheet conductance, equations in this section are expressed in terms 
of conductance $G$. 
\subsection{Semi-classical Cyclotron Resonance}

Here we demonstrate the model with which we use to check the existence of cyclotron resonance and estimate the magnitude of resonance frequency. Conductance is expressed as the sum of a Drude transport term plus a semi-classical cyclotron resonance term \cite{dresselhaus1955cyclotron}. 

\begin{equation}
G(\omega) = G_{Dr}(\omega) + G_{CCR}(\omega) 
\label{sig_classical}
\end{equation}

\begin{equation}
G_{Dr} (\omega)= G_a \frac{1}{1 - i \omega \tau_a} 
\label{sig_Dr}
\end{equation}

\begin{equation}
G_{CCR}(\omega) = G_b \frac{1 - i \omega \tau_b}{1 +(\omega_c^2-\omega^2)\tau_b^2 -2i \omega \tau_b} 
\label{sig_ccr}
\end{equation}

In this model we have five real-valued parameters, $G_a$ which is the DC conductance of the Drude transport term, $\tau_a$ which is its momentum relaxation time, $G_b$ which is the DC conductance of cyclotron resonance term, $\tau_b$ which is its relaxation time, and $\omega_c$ the resonance frequency.

\subsection{Supercyclotron Resonance} 

Davison et al. derived an expression for  a ``supercyclotron" resonance  that could originate in the hydrodynamic theory of a quantum fluctuating superconductor \cite{davison2016hydrodynamic}.  Here one can expect that one may express the conductance as a sum of a Drude transport term plus this additional term e.g.

\begin{equation}
G(\omega) = G_{Dr}(\omega) + G_{SCR}(\omega) 
\label{sig_super}
\end{equation}

with 

\begin{equation}
G_{Dr} (\omega)= G_c \frac{1}{1 - i \omega \tau_c} 
\label{sig_Dr2}
\end{equation}

\begin{equation}
G_{SCR}(\omega) = C \frac{(1-\rho_v^2)(-i\omega+\Omega)+2\rho_v\Omega_H}{(-i\omega+\Omega)^2+({\Omega^H})^2} 
\label{sig_scr}
\end{equation}

Here $G_c$, $\tau_c$, $C$, $\rho_v$, $\Omega$, and $\Omega_H$ are real and independent quantities. The frequency dependence of the $G_{SCR}$ corresponds to a hydrodynamic supercyclotron mode that oscillates at $\Omega^H$ with decay rate $\Omega$, i.e. $\omega_{\star} = \pm \Omega^H - i\Omega$ \cite{davison2016hydrodynamic}.  Although the physical meaning of the coefficients is different, not that the functional form given for the supercyclotron mode is the same as that for classical cyclotron resonance.

\subsection{Fitting Models for Fig. 2 in the Main Text}

The conductance can be fit with three terms in the superconducting phase: a delta function, a narrow Drude term to capture long-lived degrees of freedom associated with supercurrent relaxation, and a much wider Drude term that may account for a ``normal" contribution. The latter two terms remain in the anomalous metal phase.   The full expressions are

\begin{equation}
G(\omega) = G_{SC}(\omega) + G_{Dr_1}(\omega) + G_{Dr_2}(\omega),
\label{sig_tot}
\end{equation}

\begin{equation}
G_{SC}(\omega) = G_0\delta(\omega=0) + i \frac{2 G_0}{\pi \omega},
\end{equation}

\begin{equation}
G_{Dr_1} (\omega)= \frac{{\omega_{p_1}}^2}{4\pi}\frac{1}{\Gamma_1-i\omega},
\label{sig_del}
\end{equation}

\begin{equation}
G_{Dr_2}(\omega) = \frac{{\omega_{p_2}}^2}{4\pi}\frac{1}{\Gamma_2-i\omega} - i\epsilon_0(\epsilon_{\infty}-1)\omega.
\label{sig_Dr_n}
\end{equation}

In the above equations, plasma frequency $\omega_p$ is introduced as a fitting parameter and is defined as $\omega_p \equiv \sqrt{\frac{4\pi Ne^2}{m^*}}$ in a Drude model. High frequency dielectric constant in the third term accounts for interband absorptions. In practice, this fitting is done with Reffit software \cite{kuzmenko2005kramers}, and the delta function term is added by setting the scattering rate $\Gamma$ to zero (see eq. \ref*{sig_del}). Since $\epsilon_{\infty}$ always tends to be smaller than 1 in the fitting, we fix its value to be 1. In our fitting, each point in the real and imaginary part is assigned equal weight in least squared fitting. For fields above 2 T, the DC conductance is given 10 times the weight than real and imaginary AC conductance.

\begin{center}
	\begin{table*}[]
		\label{tab1}
		\caption{Fitting parameters of the two Drude terms for different magnetic fields.}
		\begin{tabular}{ | l | l| l| l| l|}
			\hline
			\diagbox{field}{parameters}
			& $\omega_{p_1}^2/4\pi$ ($\Omega^{-1} Hz$) & $\Gamma_1$ (Hz) & $\omega_{p_2}^2/4\pi$ ($\Omega^{-1} Hz$) & $\Gamma_2$  (Hz) \\ \hline
			4 T & 2.159E6 & 1.183E9 & 1.506E7 & 9.051E9 \\ \hline
			5 T & 1.215E6 & 1.607E9  & 1.624E7 & 13.27E9 \\ \hline
			6 T & 1.973E5 & 9.083E8  & 1.338E7 & 13.96E9 \\ \hline
		\end{tabular}
	\end{table*}
\end{center}

Supp. Fig. \ref{fig:Fig5} shows fitting of 4T, 5 T and 6 T data, where the two Drude contributions are plotted separately on the same graph. The fitting parameters of the Drude terms are (for convenience $\omega_p^2/4\pi $ is presented instead of $\omega_p $) collected in Table. I.  At 4 T, the magnitude of the two Drude terms are comparable in the zero frequency limit but their widths differ by almost an order of magnitude. Both terms are necessary to achieve a proper fit (particularly the imaginary part). Because the width of the "red" component (presumably related with normal charges) is larger than the upper bound of our experimental detection, the width from its fitting might not be accurate and could be larger than 9 GHz. At 5 T, and 6 T, both terms drop in magnitude but the "blue" component associated with relaxed superfluid becomes outnumbered by normal charges. 

\begin{figure}
	\centering
	\includegraphics[width=1 \textwidth]{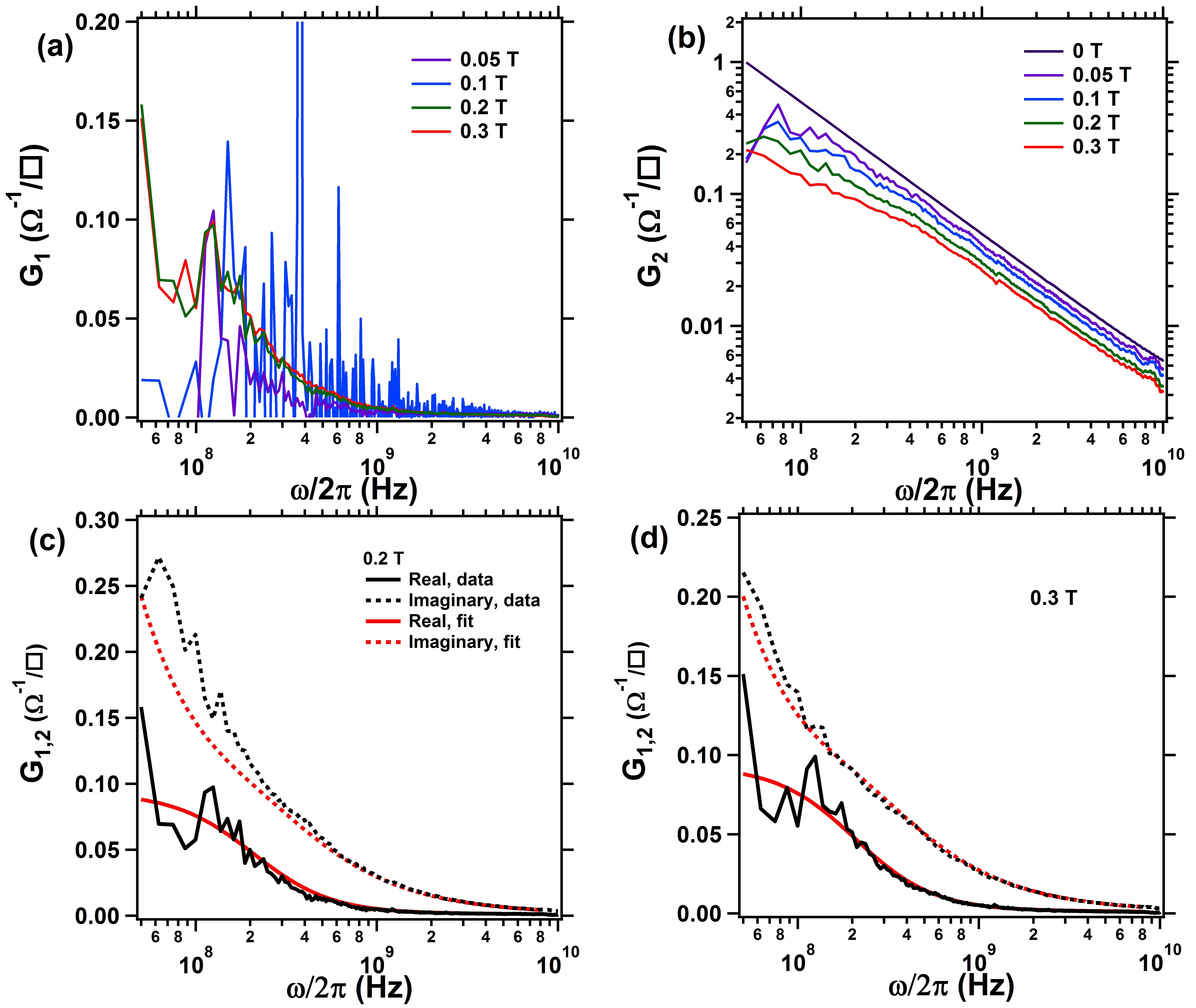}
	\caption{(color online) Complex conductance at magnetic fields smaller than 0.3 T, supplementary to Fig. 2 of the paper. (a) Real part. (b) Imaginary part. (c-d) Data and fit for 0.2 and 0.3 T with horizontal axis on log scale.}
	\label{fig:Fig6}
\end{figure}

\begin{center}
	\begin{table*}[]
		\label{tab2}
		\caption{Fitting parameters of the two Drude terms for different magnetic fields.}
		\begin{tabular}{ | l | l| l| l| l| l|}
			\hline
			\diagbox{field}{parameters}
			& $G_0$ ($\Omega^{-1} Hz$) & $\omega_{p_1}^2/4\pi$ ($\Omega^{-1} Hz$) & $\Gamma_1$ (Hz) & $\omega_{p_2}^2/4\pi$ ($\Omega^{-1} Hz$) & $\Gamma_2$  (Hz) \\ \hline
			0.2 T & 1.738007639E+07	& 1.903851280E+07 &	2.077537581E+08 & 1.873827056E+07 & 1.358669373E+10 \\ \hline
			0.3 T & 1.408368762E+07 & 1.903851244E+07 &	2.077537608E+08 & 1.873827288E+07 & 1.358669521E+10 \\ \hline
		\end{tabular}
	\end{table*}
\end{center}

\section{Small Field Data and Fit}
The conductance spectra below 0.3 T is presented in Supp. Fig. \ref{fig:Fig6} to show that there is no cyclotron resonance-like signature in the entire range we measured. The Corbino spectroscopy technique becomes less sensitive for both highly conductive (compared with the intrinsic impedance of the coaxial cables which is 50 $\omega$) and highly insulating samples\citealp{scheffler2004broadband}. At small fields the real part of conductance at finite frequencies is very small and is difficult to measure accurately. For 0.05 T, the real part conductance data is very noisy and gave a negative spectral weight calculated from integrating experimental data to 10 GHz. Therefore we did not include 0.05 T data point for Component 2 and the total spectral weight. As one can see, there is no CR feature at all small fields. The imaginary part is less noisy and we could see that it follows the qualitative trend to decrease with increasing magnetic field. The small field data is consistent with our results and conclusions. Data were fit to models mentioned in Sec.\ref{tab1}. The fitting parameters are presented in Table. II.

\begin{figure}
	\centering
	\includegraphics[width=0.45 \textwidth]{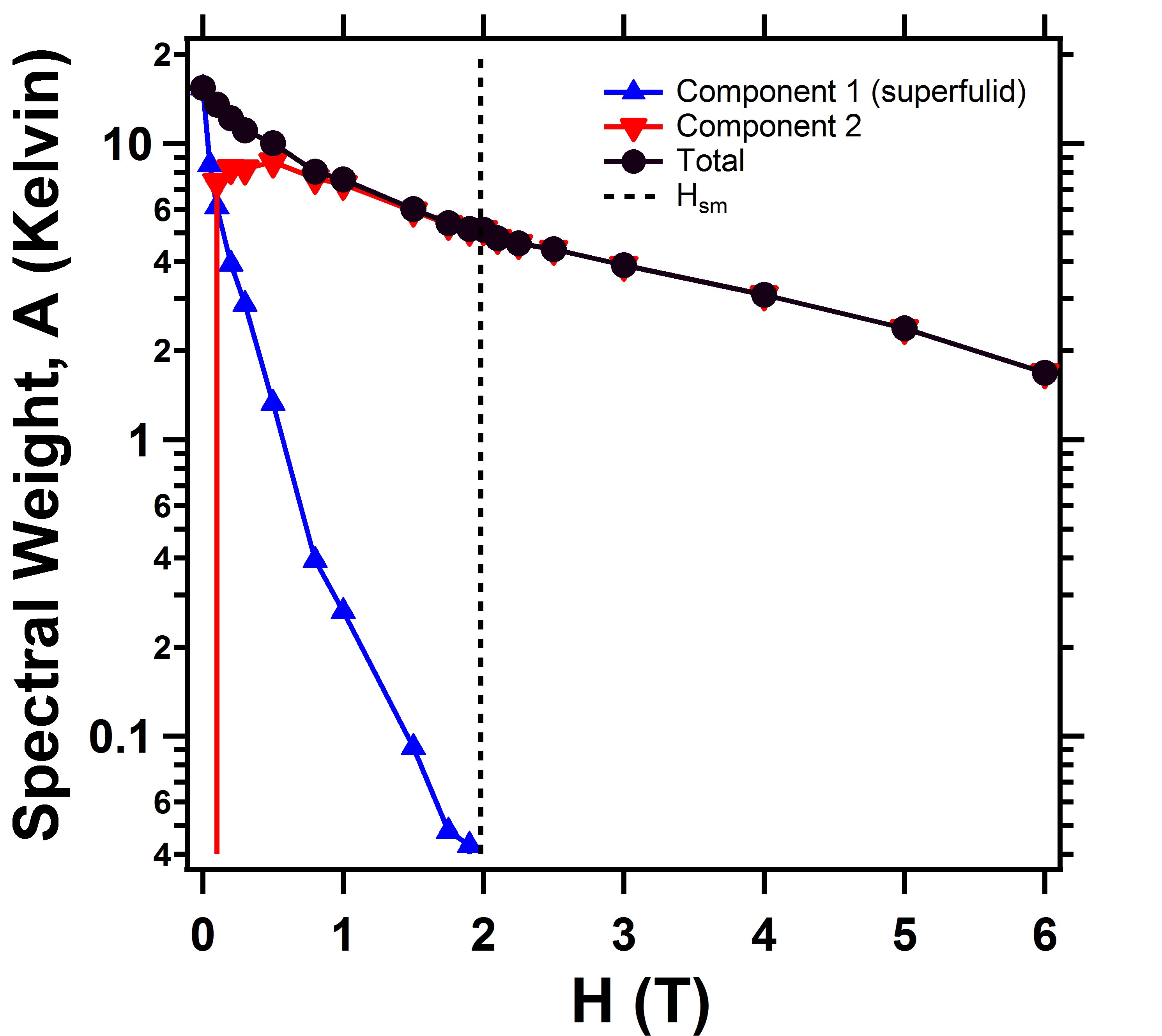}
	\caption{(color online)  Spectral weight plotted on log scale, from models fits in the units of Kelvin for the delta function and Drude terms. Blue and red corresponds to the superfluid (delta function) and the slowly decaying component respectively. The total spectral weight is in black. The vertical dashed line corresponds to our estimate of the superconductor to metal transition.   The delta function goes to zero as expected at the transition.}
	\label{fig:Fig7}
\end{figure}
\vspace{12pt} 

\section{Spectral Weight on Log Scale}
Supp. Fig. \ref{fig:Fig7} shows the same data as in Fig. 3 of the paper but with the vertical axis on log scale. It can be clearly seen in this graph that the spectral weight of Drude terms diminish much more slowly than the delta function term that corresponds to condensed Cooper pairs. 

\begin{figure}
	\centering
	\includegraphics[width=0.45 \textwidth]{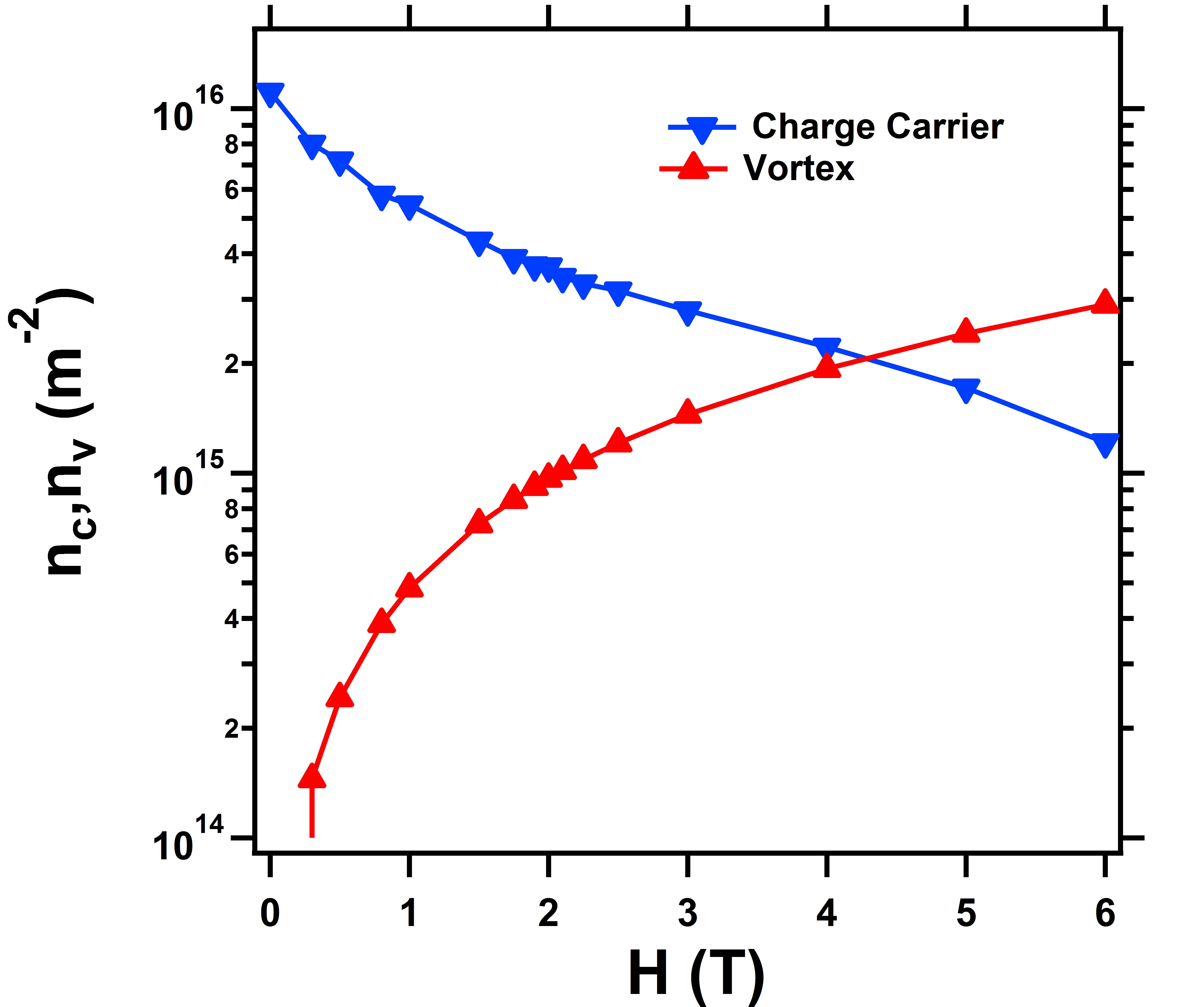}
	\caption{(color online)  Estimated charged carrier (blue) and vortex density (red) at different magnetic fields.}
	\label{fig:Fig8}
\end{figure}

\section{Estimate of Charge Carrier/Vortex Densities}
Spectra weight ($A$) in the frequency range we measured corresponds to long-lived charged degrees of freedom that contribute to transport. The ratio of the population of long-lived charges (in particular Cooper pairs) $vs.$ vortices is an important point of reference for studying field tuned quantum phase transitions, e.g. see ref. \cite{mulligan2016composite}. In this regard, we converted the spectral weight from data (integrated to 10 GHz) to charge density by:
\begin{equation}
n_c = \frac{16\pi k_B m A }{h^2}
\end{equation}
We choose to plot 1/2 of this charge density because for $H < H_{c_{2}}$ most of the charge carriers could be Cooper pairs. Because of the 10 GHz cutoff, charge density is underestimated. However, not all charged degrees of freedom are Cooper pairs. The charge density estimated herein reflects the order of magnitude of Cooper pair density. 
Vortex density is estimated using:
\begin{equation}
n_v = H/\Phi_0
\end{equation} where flux quantum $ \Phi_0 \equiv h/(2e)$. 
The result is shown in Supp. Fig. \ref{fig:Fig8}. It can be seen that near $H_{sm} \approx $ 2T the two densities are on the same order of magnitude.

\end{document}